\documentclass[aps,titlepage,12pt]{revtex4}
\usepackage{graphicx}
\usepackage{color}
\begin{document}

\title{Uncovering the information core in recommender systems}

\author{Wei Zeng$^{1,2}$}
\author{An Zeng$^{3,\ast}$}
\author{Hao Liu $^{3}$}
\author{Ming-Sheng Shang$^{1,3}$}
\author{Tao Zhou$^{1,\ast}$}
\affiliation{
$^{1}$Web Sciences Center, University of Electronic Science and Technology of China, Chengdu 611731, P.R. China\\
$^{2}$State Key Laboratory of Networking and Switching Technology, Beijing 100876, P.R. China\\
$^{3}$Department of Physics, University of Fribourg, Fribourg CH1700, Switzerland\\
$^{\ast}$ Correspondence should be addressed to an.zeng@unifr.ch, zhutou@ustc.edu
}

\begin{abstract}\noindent
\\With the rapid growth of the Internet and overwhelming amount of information that people are confronted with, recommender systems have been developed to effectively support users' decision-making process in online systems. So far, much attention has been paid to designing new recommendation algorithms and improving existent ones. However, few works considered the different contributions from different users to the performance of a recommender system. Such studies can help us improve the recommendation efficiency by excluding irrelevant users. In this paper, we argue that in each online system there exists a group of core users who carry most of the information for recommendation. With them, the recommender systems can already generate satisfactory recommendation. Our core user extraction method enables the recommender systems to achieve 90\% of the accuracy by taking only 20\% of the data into account.
\end{abstract}

\maketitle
The Internet nowadays provides us with abundant online contents, which makes it very time-consuming to go over every detail and find our needed information. This is often refereed as the information overload problem. In order to solve it, search engines and recommender systems are widely investigated \cite{Brin:1998, Adomavicius:2005:TKDE,Koren:2009:MFT,Tang:2012:KDD,Lv:PhysRep:2012}. The search engine returns the relevant contents based on the keywords given by users. Compared to the search engine, the recommender system provides more personalized services by predicting the potential interests according to users' historical choices. These techniques have already been successfully applied to some well-known web sites, such as \emph{google.com}, \emph{amazon.com}, \emph{taobao.com} and \emph{youtube.com}.

For recommendation algorithms, the most famous one from computer science is the so-called collaborative filtering (CF) with user-based and item-based versions \cite{Adomavicius:2005:TKDE, Chen:2012:SIGIR, Xu:2012:WWW}. The user-based CF estimates each user's preferences by referring to her similar users' tastes, while the item-based CF recommends items which are similar to the target user's selected items. Recently, some physical concepts have been introduced to recommendation algorithms. Since recommender systems can be naturally represented by user-object bipartite networks~\cite{PhysRevE.72.066107,huangsc531146,Shang:EPL:2010}, some classic network-based propagation processes such as mass diffusion \cite{Zhou:PRE:2007, Zhang:EPL:2007} and heat conduction \cite{Zhang:PRL:2007}, are applied to find the most relevant objects for users. The hybridization of these two propagation-based methods can effectively solve the diversity-accuracy dilemma in recommendation \cite{Zhou:PNAS:2010}. Based on these algorithms, many extensions have been made. For example, the preferential diffusion \cite{Lv:PRE:2011}, the biased heat conduction \cite{Liu:PRE:2011} and network manipulation~\cite{EPL10058005} are able to further improve the recommendation accuracy for small-degree objects (i.e. solving the cold-start problem). More recently, the long-term influence of different recommendation algorithms on network evolution has been studied~\cite{Zeng:2012:EPL}.

Related works overwhelmingly focus on designing new algorithms, while the effects of the underlying user-object bipartite networks on the recommendation results are seriously overlooked, to the best of our knowledge. More specifically, the relevance of individual users on the recommendation process has not yet been well addressed. In online systems, it is reasonable to imagine that there are some ``expert" users who know well about objects qualities in certain fields. By referring to them, the recommender systems can generate satisfactory recommendations for the user who have common interests with these expert users. Besides, there are some malicious online users who seek to bias the output of the recommender systems \cite{handbook:2011}. Eliminating these attackers is very meaningful to enhance the robustness of the recommender systems \cite{Zhou:EPL:2011}. Therefore, investigation on users' roles in recommendation can improve the efficiency as well as the robustness of recommendation algorithms by excluding irrelevance and unreliable users.

In individual level, it is already pointed out that considering $K$ most similar users to the target user can improve the recommendation accuracy under the user-based collaborative filtering framework (known as the ``KNN" method)~\cite{Adomavicius:2005:TKDE}. In this paper, we find that such phenomenon also exists in system level, i.e., one can achieve satisfactory recommendation for all users by only referring to a small group of core users. We first study the relevance of users in a recommender system and find that there exists a ``information core" consisting of some key users. The size of the core users is around $20$ percent of the whole system. The recommendation accuracy by relying only on the core users can reach $90$ percent of that with all users. This is very meaningful from practical point of view since it can significantly speed up the recommendation process in real online systems. Moveover, the analysis in this paper is helpful for the online-retailers to categorize costumers and provide better personalized services for them.

\noindent \\ \textbf{\large Results}

A recommender system can be naturally represented by a bipartite network $G(U,O,E)$, where $U=\{u_1,u_2,...,u_n\}$, $O=\{o_1,o_2,...,o_m\}$ and $E=\{e_1,e_2,...,e_l\}$ are sets of users, objects and links, respectively. The bipartite network is denoted by an adjacency matrix $A$, where the element $a_{i\alpha}=1$ if user $i$ has collected object $\alpha$, and 0 otherwise (we use Greek and Latin letters, respectively, for object- and user-related indices). The degree of an object $\alpha$ and a user $i$, $k_{\alpha}$ and $k_i$, represent respectively the number of users who have collected object $\alpha$ and the number of objects collected by user $i$. For a target user to whom we will recommend objects, each of her uncollected objects will be assigned a score by the recommendation algorithm and the top-$L$ objects with the largest scores will be recommended. Different algorithms generate different object scores and thus different recommendation lists for users.

The \emph{mass diffuse} \cite{Zhou:PRE:2007} (MD for short) algorithm works by assigning objects an initial level of ``resource" denoted by the vector $\overrightarrow{f}$ (where $f_{\alpha}$ is the resource possessed by object $\alpha$), and then redistributing it via the transformation $\overrightarrow{f'}=W\overrightarrow{f}$, where $W_{\alpha\beta}=\frac{1}{k_{\beta}}\sum_{j=1}^n\frac{a_{j\alpha}a_{j\beta}}{k_j}$ is a column-normalized $m\times m$ matrix. For a target user, the resulting recommendation list of uncollected objects is sorted according to $\overrightarrow{f'}$ in descending order and top-$L$ objects with the most resources will be recommended.

The MD method can be described in a more intuitive way: The initial resources placed on objects is first evenly divided among neighboring users and then evenly divided among those users' neighboring objects. In a real network, there can be a lot of neighboring users who have common objects with the target user. We argue here that only a few of the most relevant neighboring users should be taken into account in the diffusion. By doing this, there will be less computation in recommendation and the noisy information from the less relevant users can be reduced. Accordingly, we propose the \emph{K-Nearest Neighbor Mass Diffuse} (KNNMD) method in which only the \emph{K} nearest neighbors of the target users will be considered. Four different ways can be used to identify the most relevant neighbors: (1) random; (2) degree-based; (3) resource-based; (4) similarity-based ones. When the resources are located at the user side, the random method randomly selects $K$ users as the neighbors; the degree-based method selects $K$ users with the largest degrees as neighbors; and the resource-based method selects $K$ users with the largest received resources as the neighbors. The similarity-based method is a bit more complicated than the previous three methods. Firstly, we compute the similarities between the target user and other users. The \emph{cosine} index \cite{Manning:2008} is used to measure the similarity: $s_{ij}=|\Gamma_i\cap\Gamma_j|/\sqrt{k_ik_j}$, where $\Gamma_i$ is the set of objects being collected by user $i$. The similarity-based method selects $K$ users with highest similarities to the target user. A visual representation of KNNMD is given in Fig. 1.
\begin{figure}[!htb]
\centering
\includegraphics[width=12cm,height=8cm]{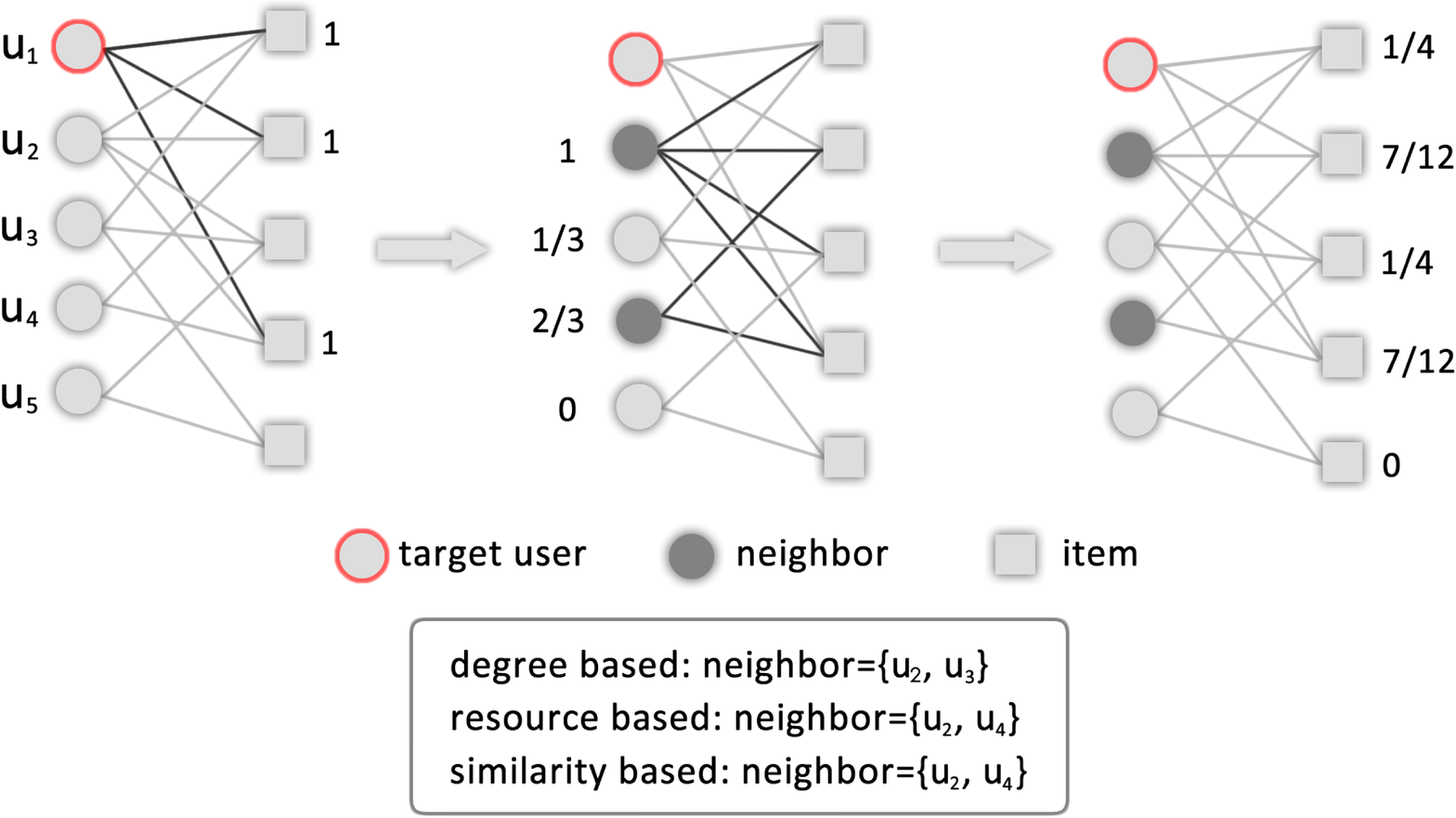}
\noindent \\ \textbf{Figure 1.} A visualization of KNNMD methods. $u_1$ is the target user and two neighbors are selected by the similarity-based method. Results from degree-based and resource-based methods are also shown.
\label{fig1:method}
\end{figure}

We compare the above four KNNMD methods on three real datasets: \emph{Douban}, \emph{Last.fm} and \emph{Flickr} (see details about the datasets in \textbf{Methods}). The metric \emph{recall} (see the definition of recall in \textbf{Methods} and the definition and results for more metrics in SI) is chosen to measure the accuracy of recommendation algorithms. A higher recall value is corresponding to a higher recommendation accuracy. The results of these KNNMD methods are presented in Fig. 2. It can be seen that the best method is the similarity-based KNNMD which outperforms the standard MD method for $K\geq20$ in Douban, $K\geq20$ in Last.fm and $K\geq40$ in Flickr, respectively. The optimal neighbor number $K^*$ of this method is around $180$ in Douban, $300$ in Last.fm and $280$ in Flickr, respectively (see Table S2 in SI). Moreover, one can see that the accuracy of the MD method is significantly improved by reducing the less relevant neighboring users (see SI for details).

\begin{figure}[!htb]
\centering
\includegraphics[width=15cm,height=6cm]{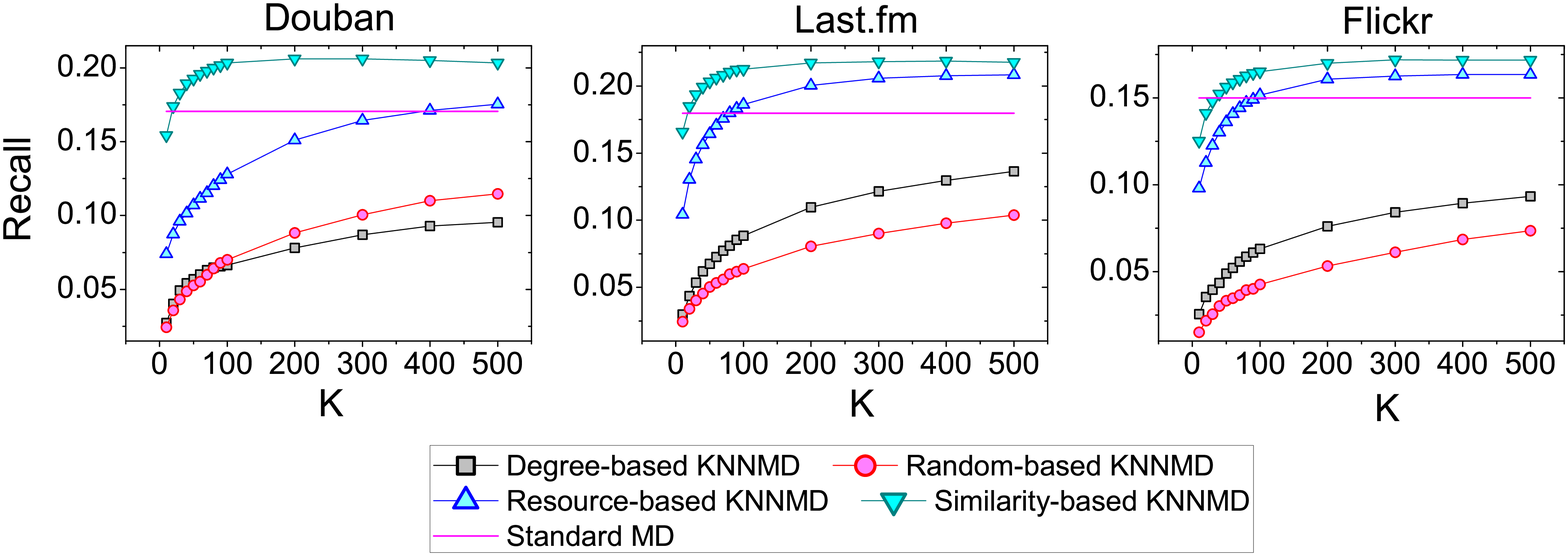}
\noindent \\ \textbf{Figure 2.} The accuracy of KNNMD methods. The recommendation length $L$ is set to $20$.
\label{fig2:knnmd_neighbor}
\end{figure}

Notice that, the above analysis is at the individual level and the selected $K$ neighbors for different individuals are different. The nice performance of KNNMD arises an important question: in the system level, which kinds of users are the most relevant ones for recommending objects for all users. We denote this group of users as the information core in the online systems.

We thus propose four approaches to assess the relevance of users and find the information core. The most straight-forward one is simply based on the degrees of users, with an underlying hypothesis that the relevance of a user can be reflected by her degree, and the information core consists of users with the largest degrees. The second one is to randomly select a set of users as the information core. This method is used as a benchmark for comparison. In the third method, we first compute the top-$N$ (e.g. $N$=10, 20, 50) most similar neighbors of each user based on the cosine similarities, and then count how many times a user has appeared in other users' top-$N$ lists. Those users appear most frequently are selected as the information core. The fourth one is similar to the third one but takes into account the ranks of each user in other users' top-$N$ neighbor lists. Supposing user $i$ belongs to user $j$'s top-$N$ neighbors and his position is $p$th, then the score of $i$ is $1/p$. If $i$ also appears in other users' top-$N$ neighbor lists, we sum his scores as his final weight: $w_i=\sum_{j, N(j)\ni i}1/p_{ij}$, where $N(j)$ is the top-$N$ neighbor set of user $j$ and $j$ runs over all users whose $N(j)$ set contains $i$. $p_{ij}$ is $i$'s position in $j$'s top-$N$ neighbor list. Finally, those users with the largest sums will be selected as the information core. A toy example of the frequency-based and the rank-based methods to find the information core from the network in Fig. 1 is illustrated in Fig. 3.

\begin{figure}[!htb]
\centering
\includegraphics[width=12cm,height=10cm]{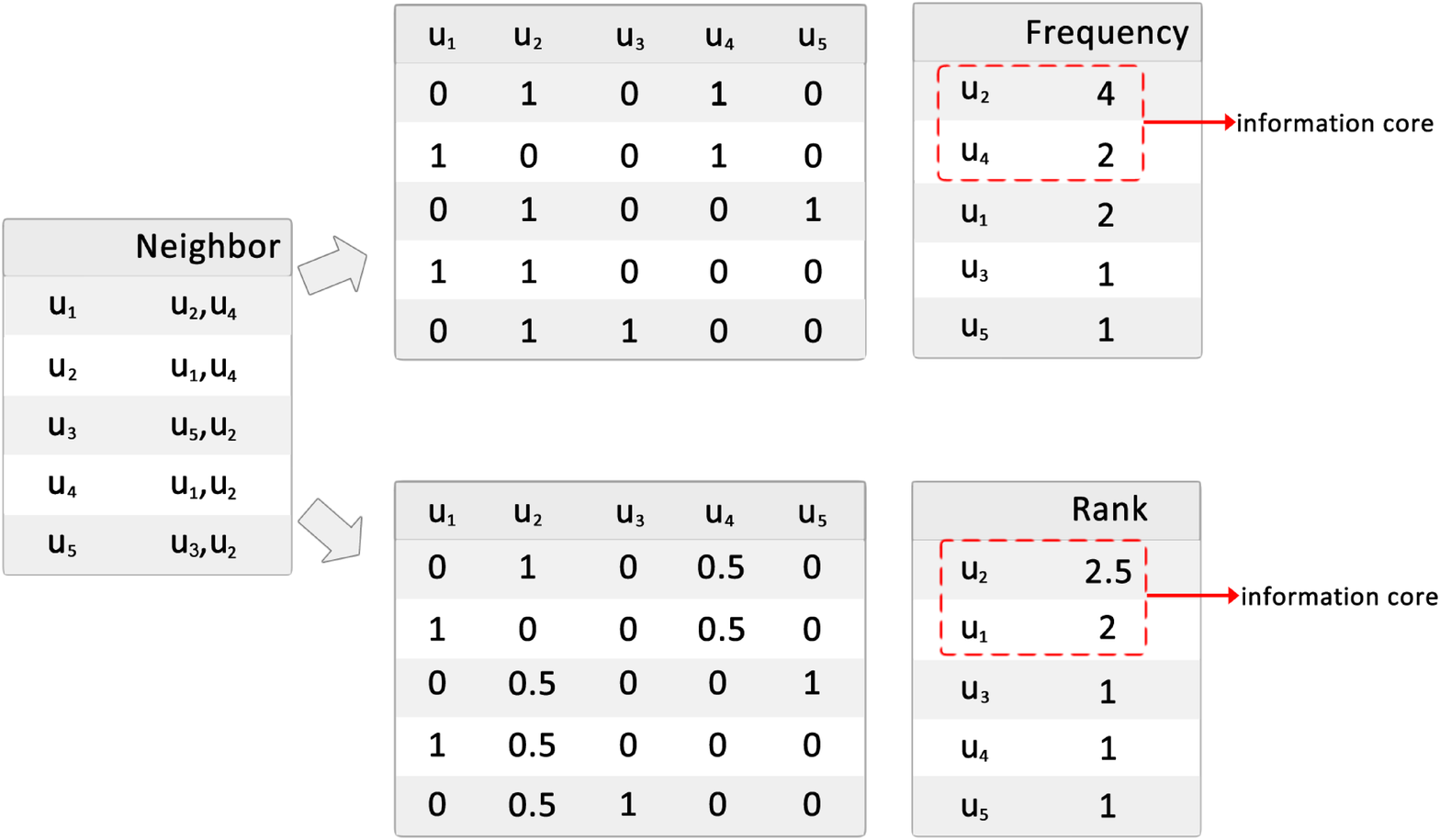}
\noindent \\ \textbf{Figure 3.} The frequency-based and rank-based method to find the information core of the user-object bipartite network in Fig. 1. For each user, we select her top-2 neighbors. The size of the information core (number of users) is set to be 2, and the information cores are $\{u_2, u_4\}$ and $\{u_2, u_1\}$ by the frequency-based and rank-based method, respectively.
\label{fig3:core_community}
\end{figure}

To study the importance of the information core in recommendation, we make use of four recommendation algorithms: MD \cite{Zhou:PRE:2007}, similarity-based KNNMD (in the following, KNNMD refers to the similarity-based KNNMD), the hybrid of the mass diffusion and heat conduction \cite{Zhou:PNAS:2010} (Hybrid for short and the details are presented in \textbf{Methods}) and user-based collaborative filtering \cite{Lv:PhysRep:2012} (UCF for short and the details are presented in \textbf{Methods}). We firstly compute the accuracy of each algorithm in the traditional way, i.e. using all users in the system. We also compute its accuracy when only the users in the information core are taken into account. Given the information core $C$ and the target user $i$, only the users in $C$ will receive the resources from $i$'s collected objects in the MD and Hybrid methods. Other users will not receive resources even though they have common objects with $i$. Then the users who have received resources redistribute the resources back the object side. For the KNNMD method, we firstly compute $i$'s top-$K$ neighbors who are in the information core $C$ and then only these $K$ neighbors will receive resources and redistribute them. Similarly, the top-$K$ neighbors will be limited in $C$ in the UCF method. This procedure is equivalent to removing non-core users from the network. However, we still make recommendations to these non-core users. The importance of the information core in recommendation can be seen by comparing the accuracy contributed by the core to that of the traditional methods. The comparison of traditional mass diffuse and the information-core-based mass diffuse is presented in Fig. 4.

\begin{figure}[!htb]
\centering
\includegraphics[width=12cm,height=10cm]{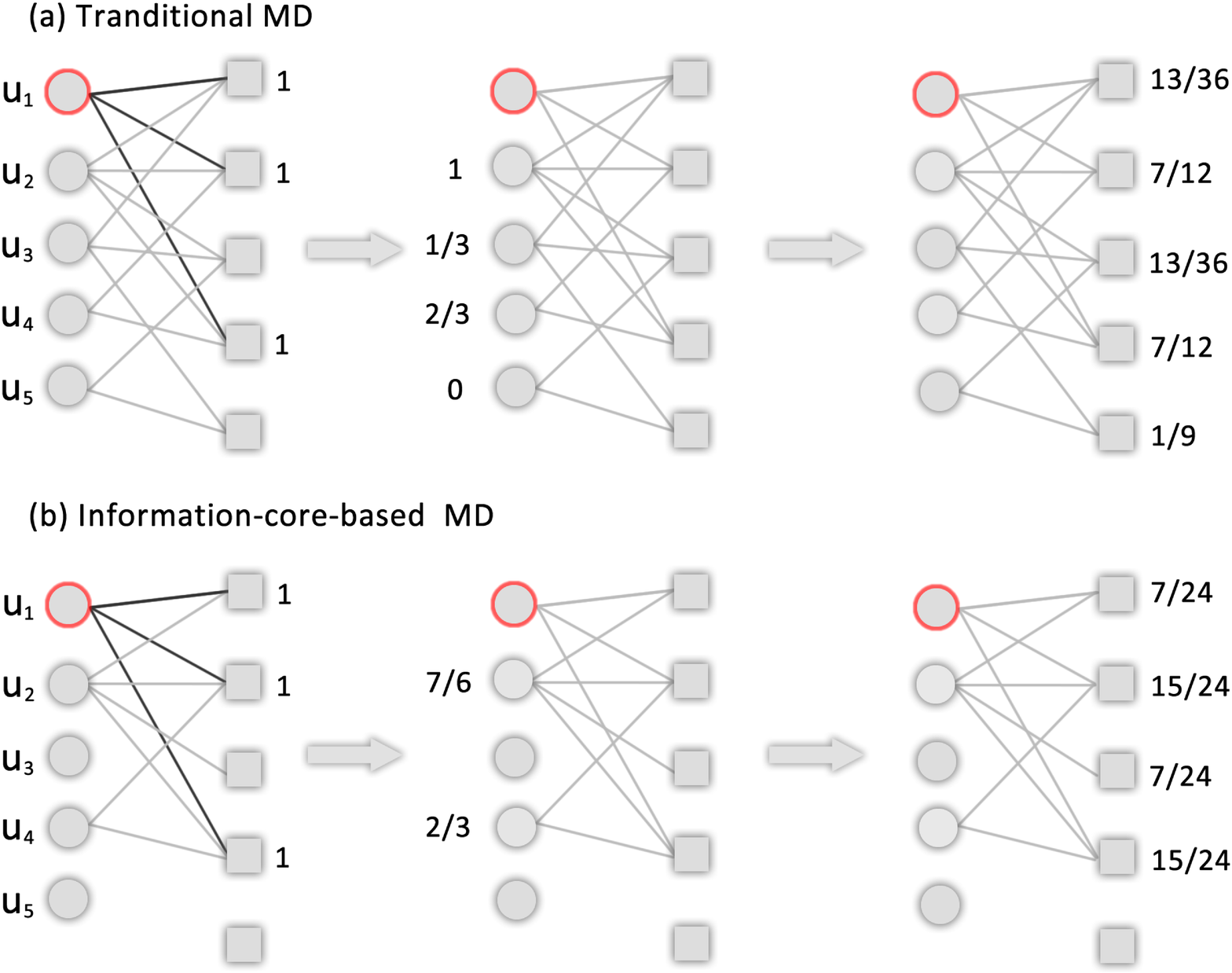}
\noindent \\ \textbf{Figure 4.} The information-core-based mass diffuse. (a) is the traditional mass diffuse method which consider all the users and (b) is the information-core-based mass diffuse which only takes into account information core users. In (b), all the irrelevant links have been removed. The information core users are $u_2$ and $u_4$ which are identified by the frequency-based method.
\label{fig2:knnmd_neighbor}
\end{figure}

We use again the recall metric to measure the accuracy of the core-based recommendation algorithms and report the results in Fig. 5 where $r$ denotes the fraction of users in the information core. When $r=1$, all the users will be used in the recommendation algorithms, equivalent to the traditional method. Generally speaking, the recommendation accuracy tends to decrease with $r$ since the available information for the recommendation algorithm is less. The accuracy decreases slower when we choose the rank-based method to identify the information core. Taking the KNNMD method for the Douban data for example, $91.4\%$ ($0.1886/0.2063$) of the accuracy can be achieved when we only use $20\%$ of users ($r=0.2$). Specifically, for the MD method in the Douban Data, the accuracy with only $20\%$ users ($r=0.2$) can be even slightly better than that with all users ($r=1$). Similar results are also observed in the other two datasets. This is of great importance since the algorithmic efficiency of recommendation methods can be largely improved if we consider fewer users.

\begin{figure}[!htb]
\centering
\includegraphics[width=15cm,height=15cm]{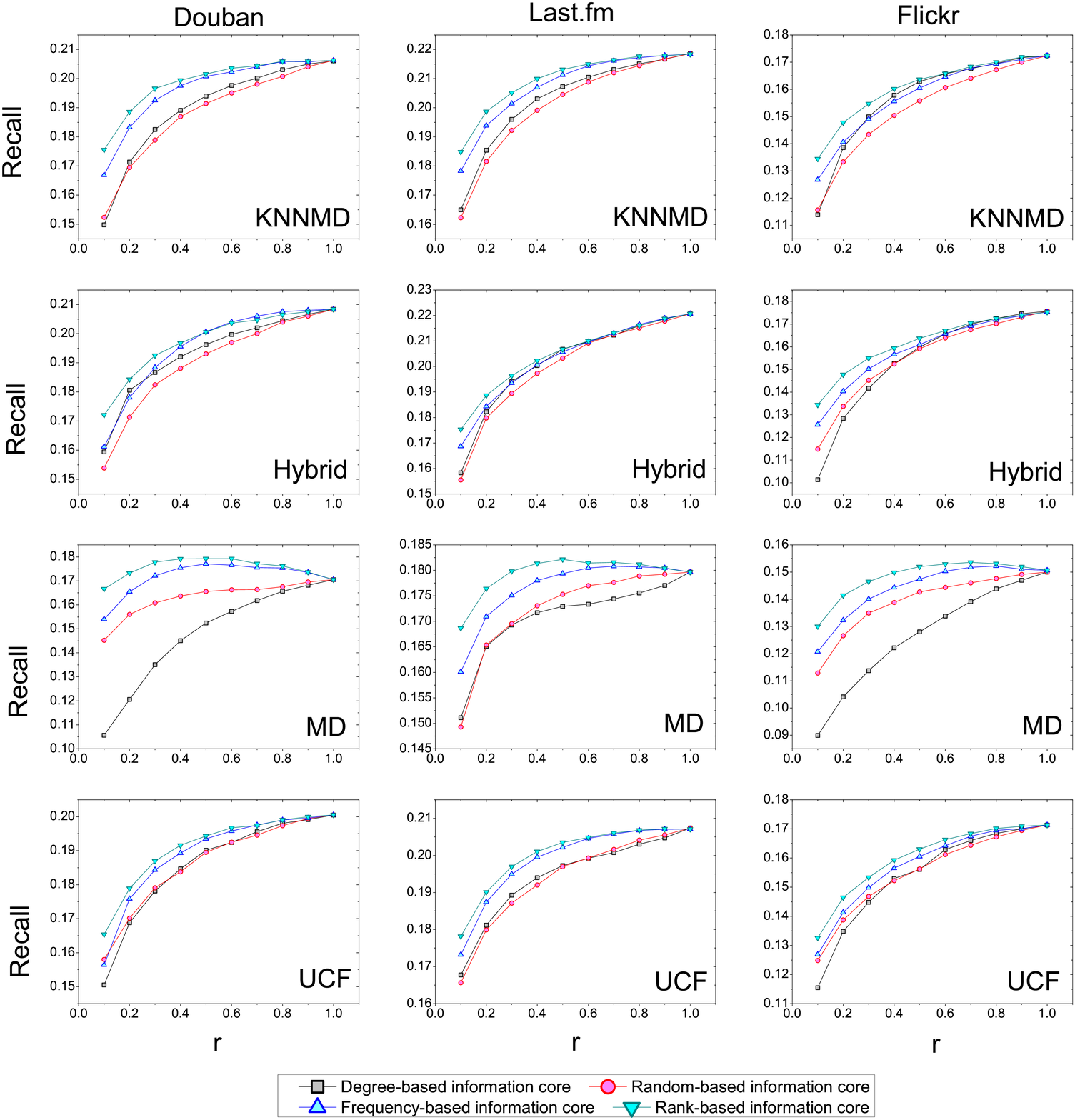}
\noindent \\ \textbf{Figure 5.} The recommendation accuracy contributed only by the information core in the recommender systems. The recommendation length $L$ is set to $20$. For the frequency-based and rank-based method, we select each user's top-20 neighbors. $r$ is the ratio of the size of the information core to the whole system.
\label{fig4:core_community_accuracy}
\end{figure}


From the above results, it can be seen that the rank-based method is better than the frequency-based method in identifying the information core, which indicates that the rank of a user appearing in others' top-$N$ neighbor list matters when assessing her relevance in the recommender system. If a user appears in most users' top-$N$ neighbor list with high rank, she should be considered as the key member in the online system since many users' recommendation will rely on her. Both methods are generally better than the random and degree-based methods. Among these methods, the degree-based method is the worst, which indicates that the large-degree users are not for sure the ``expert". Taking the MD method in the Douban data for instance, the accuracy of the degree-based method is much lower than that of the rank-based method when $r=0.2$. In many previous works about real networks with heterogenous degree distribution, attention has been overwhelmingly paid to the hubs (nodes with largest degree). Our finding here suggests that degree may not be the proper criterion to judge the importance of nodes in the information filtering process, perspectively analogous to the week ties effects in information filtering \cite{LV:EPL:2010}.

\noindent \\ \textbf{\large Discussion}

\noindent During the past decade, recommender systems have been widely investigated in several research fields, including computer science, physics, sociology and so on \cite{Lv:PhysRep:2012}. Up to now, a lot of recommendation algorithms have been proposed. However, little attention was paid to studying the effect of the underlying user-object bipartite network on recommendation process. In this paper, we study the relevance of individual users and find that there exists a information core whose size is small compared to the whole network. The users in the information core usually appear in many users' top-$N$ neighbor lists with high ranks. For many recommendation algorithms, one can achieve very good recommendation accuracy by only using the core users. Actually, similar idea can be extended to the item-based collaborative filtering. One can use only the links of those core users to calculate the items' similarity matrix and obtain accurate recommendation~\cite{Blattner2007753}. This work may find wide applications in practice. For one thing, it can significantly speed up the recommendation process in real online systems since the recommendation engine only has to deal with a small fraction of data. For another, the analysis in this paper can be also helpful for the online-retailers to categorize costumers and provide better personalized services for them.

There are still many open issues, such as extending similar technique to monopartite networks for link prediction \cite{Lv:Physica:2011}. Another interesting open issue is to study the location of these core users in the network. Specifically, one can investigate whether the core users are diversely distributed in different communities. Related study may lead to some better topology-based method to identify the core users in networks. Finally, the evolution of the information core is also an important topic. A relatively stable information core over time will lower the frequency to update core users, and thus further reduce the computational cost in practice.

\noindent \\ \textbf{\large Methods}

\noindent  \textbf{Data description.} We use three datasets to test the accuracy of algorithms, namely Douban \cite{Huang:2012:WSDM}, Last.fm \cite{Celma:Springer2010} and Flickr \cite{Mislove:2007:MAO}. Douban (www.douban.com), launched on March 6, 2005, is a Chinese Web 2.0 web site providing user rating, review and recommendation services for movies, books and music. It is also the largest online Chinese book, movie and music database and one of the largest online communities in China. The raw data contains user activities before Aug 2010 and we randomly sample 17,000 users who have collected at least ten songs. The Last.fm (www.last.fm) is a worldwide popular social music site and the objects in this dataset are referred to the artists which can be collected from Last.fm API. The raw data consists of 360,000 users and we randomly sample 30,000 users who have collected at least five items (artists). Flickr (www.flickr.com) is a photo-sharing site based on a social network. The data used in this paper is individuals' group membership in Flickr, which refers to the their participation in groups. Accordingly, we provide group recommendations for users instead of objects \cite{Zeng:2013:HDF,Chen:2013:ESWA}. We randomly sample 30,000 users who have joined at least ten groups. We treat the user-object (user-group) interaction matrix as binary, that is, the element equals to 1 if the user has viewed or rated the object (joined the group) and 0 otherwise (see Table 1).

\begin{table}[!htb]
\centering
\noindent \\ \textbf{Table 1.} The statistics of Douban, Last.fm and Flickr datasets. The sparsity is defined as $\frac{l}{n\times m}$.
\label{tabS1:data_statistics}
\begin{tabular}{l|c|c|c|c}
\hline \hline
     Dataset & \#users, n & \#objects, m & \#links, l & sparsity \\[2pt]
    \hline
    Douban & 17,000 & 223,823 & 2,109,749 & $5.54\times 10^{-4}$  \\[2pt]
    Last.fm & 30,000 & 87,082 & 1,467,235 & $5.62\times 10^{-4}$ \\[2pt]
    Flickr & 30,000 & 61,352 & 1,924,461 & $7.37\times 10^{-4}$ \\[2pt]
    \hline\hline
\end{tabular}
\end{table}

\noindent  \textbf{Evaluating recommender systems.} Each data is randomly divided into two parts: the training set $E^T$ and the probe set $E^P$. The training set contains 80\% of the original links and the recommendation algorithm runs on it \cite{Jamali:2010:RS}. The rest of the links forms the probe set, which will be used to assess the performance of the recommendation algorithm. The result is obtained by averaging over five runs with independently random division of training set and probe set \cite{Lv:PRE:2011}.

For each user $i$, she may have certain number of links (corresponding to objects) in the probe set, we denote it as $E_i$. After the recommendation list (with length $L$) is generated for user $i$, we will calculate $d_i(L)$ as the number of her objects in the probe set which appear in the recommendation list. The recall of this user is defined as $R_i(L) = d_i(L)/E_i$ and the recall of the whole system is defined as $R(L) = \frac{1}{n}\sum_{i=1}^n{R_i(L)}$. A higher recall value indicates a higher accuracy of the recommendation algorithm \cite{Herlocker:2004:ACMTrans}.

\noindent  \textbf{Hybrid algorithm} When recommending objects for user $i$, the hybrid method works by assigning each object collected by user $i$ one unit of resource. The initial resources are denoted by the vector $\overrightarrow{f}$ where $f_{\alpha}$ is the resource possessed by object $\alpha$. Then they will be redistributed via the transformation $\overrightarrow{f'}=W\overrightarrow{f}$, where $W_{\alpha\beta}=\frac{1}{k_{\alpha}^{1-\lambda}k_{\beta}^{\lambda}}\sum_{j=1}^{n}\frac{a_{j\alpha}a_{j\beta}}{k_j}$
is the redistribution matrix, with $k_{\alpha}=\sum_{l=1}^{n}a_{l\alpha}$ and $k_j=\sum_{\gamma=1}^{m}a_{j\gamma}$ denoting the degree of object $\alpha$ and user $j$, respectively. $\lambda$ is a tunable parameter which adjusts the relative weight between the mass diffusion algorithm ($\lambda=1$) and heat conduction algorithm ($\lambda=0$) \cite{Zhou:PNAS:2010}.

\noindent  \textbf{User-based collaborative filtering.} In the user-based collaborative filtering method, the basic assumption is that similar users usually collect the same objects. Accordingly, the recommendation score of object $\alpha$ for the target user $i$ is $p_{i\alpha} = \sum_{j\in N(i)}{s_{ij}a_{j\alpha}}$, where $N(i)$ is the top-K neighbors of user $i$ and $s_{ij}$ is their similarity. The cosine index is chosen to measure their similarity: $s_{ij} = \sum_{\alpha=1}^m{\frac{a_{i\alpha}a_{j\alpha}}{\sqrt{k_ik_j}}}$, where $k_i$ is the degree of user $i$.

\bibliography{refs}

\clearpage

\noindent  \textbf{Acknowledgments} \\
This work is supported by the National Natural Science Foundation of China (Grant Nos. 61370150 and 91324002), the Open Foundation of State key Laboratory of Networking and Switching Technology (SKLNST-2013-1-18) and the Special Project of Sichuan Youth Science and Technology Innovation Research Team (No.2013TD0006). W.Z. acknowledges the support from the Program of Outstanding PhD Candidate in Academic Research by UESTC (YBXSZC20131029).

\noindent \\ \textbf{Author contributions} \\
WZ, AZ, HL and MSS designed the research. WZ performed the experiments, WZ, AZ, HL and TZ analysed the data, WZ, AZ and TZ wrote the manuscript.

\noindent \\ \textbf{Additional information} \\
\textbf{ Supplementary Information} for ``Uncovering information core in recommender systems".

\noindent \textbf{Competing financial interests:} The authors declare no competing financial interests.

\end{document}